\documentclass[onecolumn,showpacs]{revtex4}
\usepackage{amssymb,amsmath}
\usepackage{graphicx}
\newcommand{\aut}[2]{{#2.\ #1}}
\newcommand{\refs}[6]{#2, {\bf #3} {#4} (#5)}

\newcommand{\amp}{and }

\newcommand{\bn}{\hat{\mathbf{n}}}

\newcommand{\bm}{\mathbf{m}}

\newcommand{\beq}{\begin{equation}}
\newcommand{\eeq}{\end{equation}}
\newcommand{\bea}{\begin{eqnarray}}
\newcommand{\eea}{\end{eqnarray}}
\newcommand{\ba}{\begin{array}}
\newcommand{\ea}{\end{array}}
\newcommand{\rad}{r}    
\newcommand{\da}{d_A}
\newcommand{\pp}{{\phi}}
\newcommand{\lens}{{\rm len}}

\newcommand{\ApJ}{Astrophys. J.}
\newcommand{\PRL}{Phys. Rev. Lett.}
\newcommand{\PRD}{Phys. Rev. D}
\newcommand{\MNRAS}{Mon. Not. R. Astron. Soc.}

\newlength{\sizeonefig}
\newlength{\sizetwofig}
\begin{document}

\title{Non-Gaussian Covariance of CMB $B$-modes of Polarization and Parameter Degradation}

\author{Chao Li$^1$, Tristan L. Smith$^1$, and Asantha Cooray$^2$}
\affiliation{$^1$ Theoretical Astrophysics, California Institute
of Technology, Mail Code 103-33 Pasadena, California 91125 \\
$^2$Center for Cosmology, Department of Physics and Astronomy, 4129  Frederick
Reines Hall, University of California, Irvine, CA 92697}

\date{\today}

\begin{abstract}
The $B$-mode polarization lensing signal is a useful probe of the
neutrino mass and to a lesser extent the dark energy equation of state as the signal
depends on the integrated mass power spectrum between us and the
last scattering surface. This lensing $B$-mode signal, however, is
non-Gaussian and the resulting non-Gaussian covariance to the
power spectrum cannot be ignored as correlations between $B$-mode
bins are at a level of 0.1. For temperature and $E$-mode
polarization power spectra, the non-Gasussian covariance is not
significant, where we find correlations at the $10^{-5}$
level even for adjacent bins. The resulting degradation on neutrino mass and dark energy
equation of state is about a factor of 2 to 3 when compared to the
case where statistics are simply considered to be Gaussian. We
also discuss parameter uncertainties achievable in upcoming
experiments and show that at a given angular resolution for
polarization observations, increasing the sensitivity beyond a
certain noise value does not lead to an improved measurement of the
neutrino mass and dark energy equation of state with $B$-mode power
spectrum. For Planck, the resulting constraints on the sum of the
neutrino masses is $\sigma_{\Sigma m_\nu} \sim 0.2\ \mathrm{eV}$ and on the dark energy
equation of state parameter we find, $\sigma_w \sim 0.5$.
\end{abstract}


\maketitle

\section{Introduction}

The applications of cosmic microwave background (CMB) anisotropy
measurements are well known \cite{Kno95}; its ability to constrain
most, or certain combinations of, parameters that define the
currently favorable cold dark matter cosmologies with a
cosmological constant is well demonstrated with anisotropy data
from Wilkinson Microwave Anisotropy Probe \cite{Sper06}.
Furthermore the advent of high sensitivity CMB polarization
experiments with increasing sensitivity \cite{keating} suggests
that we will soon detect the small amplitude $B$-mode polarization
signal. While at degree scales one expects a unique $B$-mode
polarization signal due to primordial gravitational waves
\cite{KamKosSte97}, at arcminute angular scales the dominant
signal will be related to cosmic shear conversion of $E$-modes to
$B$-modes by the large-scale structure during the photon propagation
from the last scattering surface to the observer today
\cite{Zaldarriaga:1998ar}.

This weak lensing of cosmic microwave background (CMB)
polarization by intervening mass fluctuations is now well studied in the literature
\cite{lensing,Hu00}, with a significant effort spent on improving
the accuracy of analytical and numerical calculations (see, recent review in Ref.~\cite{Cha}).
As discussed in recent literature \cite{manoj},  the lensing $B$-mode signal carries important cosmological information
 on the neutrino mass and possibly the dark energy, such as its equation of state \cite{manoj}, as the lensing signal depends on the integrated mass power spectrum between us
and the last scattering surface, weighted by the lensing kernel.
The dark energy dependence involves the angular diameter
distance projections while the effects related to a non-zero neutrino mass come  from suppression of
small scale power below the free-streaming scale.

Since the CMB lensing effect is inherently a non-linear process,
the lensing corrections to CMB temperature and polarization are
expected to be highly non-Gaussian. This non-Gaussianity at the
four-point and higher levels are exploited when reconstructing the
integrated mass field via a lensing analysis of CMB temperature
and polarization \cite{HuOka02}. The four point correlations are
of special interest since they also quantify the sample variance
and covariance of two point correlation or power spectrum
measurements \cite{Scoetal99}. A discussion of lensing covariance
of the temperature anisotropy power spectrum is available in
Ref.~\cite{Coo02}. In the case of CMB polarization, the existence
of a large sample variance for $B$-modes of polarization is already
known \cite{smith}, though the effect on cosmological parameter
measurements is yet to be quantified. Various estimates on
parameter measurements in the literature ignore the effect of
non-Gaussianities and could have overestimated the use of CMB
$B$-modes to tightly constrain parameters such as a neutrino mass or
the dark energy equation of state. To properly understand the
extent to which future polarization measurements can constrain
these parameters, a proper understanding of non-Gaussian
covariance is needed.

Here, we discuss the temperature and polarization covariances due to gravitational lensing. Initial calculations on this
topic are available in Refs.~\cite{smith,smith2}, while detailed calculations on the CMB lensing trispectra
are in Ref.~\cite{Hu01}. Here, we focus mainly on the covariance and calculate them under the exact all-sky formulation;
for flat-sky expressions of the trispectrum, we refer the reader to Ref.~\cite{HuOka02}.
We extend those calculations and also discuss the impact on cosmological parameter estimates. This paper is organized as follows:
In \S \ref{calculation}, we introduce the basic ingredients for the present calculation and present
covariances of temperature and polarization spectra. We discuss our results in
\S \ref{sec:results} and  conclude with a summary in \S \ref{sec:summary}.

\section{Calculational Method}
\label{calculation}

The lensing of the CMB is a remapping of temperature and polarization anisotropies
by gravitational angular deflections during the propagation.
Since lensing leads to a redistribution of photons,
the resulting effect appears only at second order \cite{Cha}.
In weak gravitational lensing, the deflection angle on the sky is
given by the angular gradient of the lensing
potential, $\delta(\bn) = \nabla \phi(\bn)$,
which is itself a projection of the gravitational
potential $\Phi$:
\begin{eqnarray}
\phi(\bm)
&=&
- 2 \int_0^{\rad_0} d\rad \frac{\da(\rad_0-\rad)}{\da(\rad)\da(\rad_0)}
                \Phi (\rad,\hat{{\bf m}}\rad ) \,,
\label{eqn:lenspotential}
\end{eqnarray}
where $r(z)$ is the comoving distance along the line of sight, $r_0$ is the comoving distance to the surface of last scattering, and $\da(r)$ is the angular diameter distance. Taking the multipole moments,
the power spectrum of lensing potentials is now given through
\begin{equation}
        \left< \phi_{l m}^* \phi_{l m} \right>  = \delta_{l l'} \delta_{m m'} C_l^\pp
\end{equation}
as
\begin{equation}
C_l^{\phi} = \frac{2}{\pi} \int k^2\, dk P(k)
                I_l^\lens(k) I_l^\lens(k) \,,
\label{eqn:cllens}
\end{equation}
where
\begin{eqnarray}
I_l^{\rm len}(k)& =&
                \int_0^{\rad_0} d\rad W^{\rm len}(k,r)
                 j_l(k\rad)  \,,\nonumber\\
W^{\rm len}(k,r)& =&
                -3 \Omega_m \left({H_0 \over k}\right)^2
                F(r) {\da(\rad_0 - \rad) \over
                \da(\rad)\da(\rad_0)}\,,
\label{eqn:lensint}
\end{eqnarray}
where $F(r)=G(r)/a(r)$ and $G(r)$ is the growth factor, which describes the growth of
large-scale density perturbations. In our calculations
we will generate $C_l^{\phi\phi}$ based on a non-linear
description of the matter power spectrum $P(k)$. In the next three
subsections we briefly outline the power spectrum covariances
under gravitational lensing for temperature and polarization $E$-
and $B$-modes. In the numerical calculations described later, we
take a fiducial flat-$\Lambda$CDM cosmological model with
$\Omega_b=0.0418$, $\Omega_m=0.24$, $h=0.73$, $\tau=0.092$,
$n_s=0.958$, $A(k_0 = 0.05\ \mathrm{Mpc}^{-1})=2.3\times10^{-9}$, $m_\nu=0.05$ eV,  and $w=-1$. This
model is consistent with recent measurements from WMAP
\cite{Sper06}.

\subsection{Temperature anisotropy covariance}

The trispectrum for the unlensed temperature can be written in terms of the multipole moments of the temperature $\theta_{lm}$ as
\cite{Hu01}
\bea \langle
\theta_{l_1m_1}\theta_{l_2m_2}\theta_{l_3m_3}\theta_{l_4m_4}
\rangle&=&C_{l_1}^\theta C_{l_4}^\theta
(-1)^{m_1+m_4}\delta_{l_1l_3}^{m_1-m_3}\delta_{l_2l_4}^{m_2-m_4}+C_{l_1}^\theta
C_{l_3}^\theta
(-1)^{m_1+m_3}\delta_{l_1l_2}^{m_1-m_2}\delta_{l_3l_4}^{m_3-m_4}\nonumber\\
&+& C_{l_1}^\theta C_{l_2}^\theta
(-1)^{m_1+m_2}\delta_{l_1l_4}^{m_1-m_4}\delta_{l_2l_3}^{m_2-m_3}
\, . \eea It is straight forward to derive the following expression for the
multipole moment of lensed $\theta$ field as a perturbative
equation related to the deflection angle \cite{Hu00}: \bea
\tilde{\theta}_{lm}=\theta_{lm}+
\sum_{l_1m_1l_2m_2}\phi_{l_1m_1}\theta_{l_2m_2}I_{ll_1l_2}^{mm_1m_2}
+\frac{1}{2}\sum_{l_1m_1l_2m_2l_3m_3}\phi_{l_1m_1}\theta_{l_2m_2}\phi_{l_3m_3}^*
J_{ll_1l_2l_3}^{mm_1m_2m_3} \, , \label{lensedT}\eea where the
mode-coupling integrals  between the temperature field and the
deflection field, $I_{ll_1l_2}^{mm_1m_2}$ and
$J_{ll_1l_2l_3}^{mm_1m_2m_3}$, are defined in \cite{Hu01,CooLi}.

As for the covariance of the temperature anisotropy powerspectrum,
we write \bea \text{Cov}_{\theta\theta} \equiv
\frac{1}{2l_1+1}\frac{1}{2l_2+1}\sum_{m_1m_2}\langle
\tilde{\theta}_{l_1m_1}\tilde{\theta}_{l_1m_1}^*\tilde{\theta}_{l_2m_2}\tilde{\theta}_{l_2m_2}^*
\rangle
-\tilde{C}_{l_1}^\theta\tilde{C}_{l_2}^\theta=\mathcal{O}+\mathcal{P}
+(\mathcal{Q}+\mathcal{R})\delta_{l_1l_2}\label{thetacov}\eea
where the individual terms are \bea
\mathcal{O}&=&\frac{2}{(2l_1+1)(2l_2+1)}\sum_{L}{C_L^\phi}\left[(F_{l_1Ll_2}C_{l_2}^\theta)^2
+(F_{l_2Ll_1}C_{l_1}^\theta)^2\right]
\nonumber\\
\mathcal{P}&=&\frac{4}{(2l_1+1)(2l_2+1)}\sum_{L}C_L^\phi
C_{l_1}^\theta
C_{l_2}^\theta F_{l_1Ll_2} F_{l_2Ll_1} \nonumber \\
\mathcal{Q}&=&\frac{4}{(2l_1+1)^2}\sum_{L,l'}{C_L^\phi}C_{l'}^\theta
C_{l_1}^\theta
(F_{l_1Ll'})^2\nonumber\\
\mathcal{R}&=&-\frac{(l_1(l_1+1))}{2\pi(2l_1+1)}\sum_LC_L^\phi(C_{l_1}^\theta)^2
L(L+1)(2L+1) \, , \eea and the last two terms, which are related
to the Gaussian variance, can be written in terms of the lensed
temperature anisotropy power spectrum as \bea
\mathcal{Q}+\mathcal{R}=\frac{2}{2l_1+1}(\tilde{C}_{l_1}^\theta)^2
\, , \eea where \bea \tilde{C}_l^\theta&=&[1-(l^2+l)R]C_{l}^\theta+
\sum_{l_1l_2}C_{l_1}^\phi\frac{(F_{ll_1l_2})^2}{2l+1}
C_{l_2}^\theta\nonumber\\
R&=&\frac{1}{8\pi}\sum_{l_1}l_1(l_1+1)(2l_1+1)C_{l_1}^\phi\nonumber\\
F_{ll_1l_2}&=&\frac{1}{2}[l_1(l_1+1)+l_2(l_2+1)-l(l+1)]\sqrt
{\frac{(2l+1)(2l_1+1)(2l_2+1)}{4\pi}}\begin{pmatrix}l&l_1&l_2\\0&0&0\end{pmatrix} \label{eq:temp}.\eea
We note that Eqs.~(\ref{eq:temp}) are readily derivable when considering the lensing effect on the temperature anistropy spectrum as in Ref.~\cite{Hu00}.
\subsection{$E$-mode Polarization Covariance}

Similar to the case with temperature,  the trispectrum for an
unlensed $E$-field can be written in terms of the multipole
moments of the $E$-mode $E_{lm}$:
 \bea
\langle E_{l_1m_1}E_{l_2m_2}E_{l_3m_3}E_{l_4m_4} \rangle&=&C_{l_1}^E
C_{l_4}^E
(-1)^{m_1+m_4}\delta_{l_1l_3}^{m_1-m_3}\delta_{l_2l_4}^{m_2-m_4}+C_{l_1}^E
C_{l_3}^E
(-1)^{m_1+m_3}\delta_{l_1l_2}^{m_1-m_2}\delta_{l_3l_4}^{m_3-m_4}\nonumber\\
&+& C_{l_1}^E C_{l_2}^E
(-1)^{m_1+m_2}\delta_{l_1l_4}^{m_1-m_4}\delta_{l_2l_3}^{m_2-m_3}
\, \label{unlensedetris} . \eea
\begin{figure*}[t]
\includegraphics[trim = 32 1 0 0,scale=0.35,angle=0]{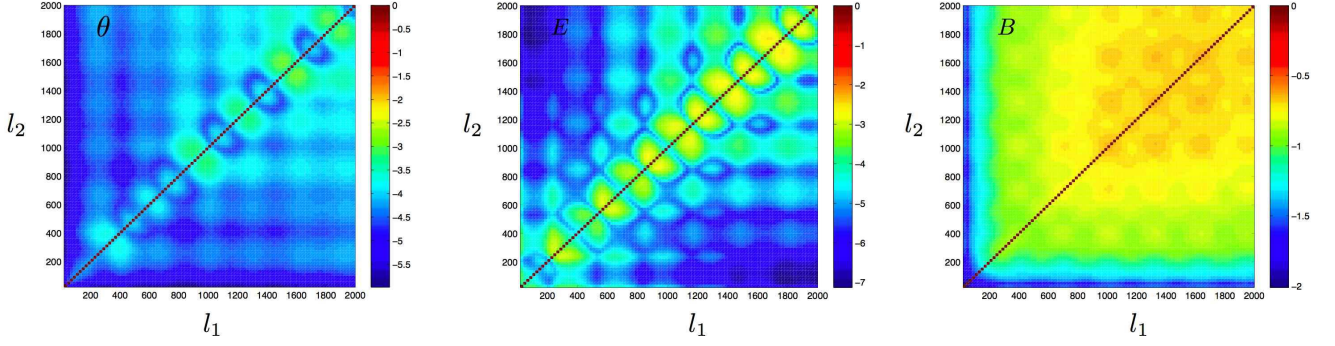}
\caption{The correlation matrix [Eq.~(\ref{eq:correlation})] for temerature (left), $E$-mode (middle), and $B$-mode (right) power spectra
between different $l$ values. The color axis is on a log scale and
each scale is different for each panel. As is clear from this
figure, the off diagonal correlation is weak for both $\theta$ and $E$-mode
power spectra, but is more than $0.1$ for most entries for the $B$-mode
power spectrum.  This clearly shows that the non-Gaussianities are
most pronounced for the $B$-mode signal and will impact the information
extraction from the angular power spectrum of $B$-modes than under
the Gaussian variance alone.  The $B$-mode covariance shown in the left panel agrees with Figure 5 of Ref.~\cite{smith2}.} \label{density}
\end{figure*}
To complete the calculation, besides the trispectrum of the
unlensed $E$-field in Eq.~\eqref{unlensedetris}, we also require
the expression for the trispectrum of the lensing potentials.
Under the Gaussian hypothesis for the primordial $E$-modes  and
ignoring non-Gaussian corrections to the $\phi$ field, the lensing
trispectra is given by \bea
\langle\phi_{l_1m_1}\phi_{l_2m_2}\phi_{l_3m_3}\phi_{l_4m_4}
\rangle=C_{l_1}^\phi C_{l_4}^\phi
(-1)^{m_1+m_4}\delta_{l_1l_3}^{m_1-m_3}\delta_{l_2l_4}^{m_2-m_4}+C_{l_1}^\phi
C_{l_3}^\phi
(-1)^{m_1+m_3}\delta_{l_1l_2}^{m_1-m_2}\delta_{l_3l_4}^{m_3-m_4}\nonumber\\+
C_{l_1}^\phi C_{l_2}^\phi
(-1)^{m_1+m_2}\delta_{l_1l_4}^{m_1-m_4}\delta_{l_2l_3}^{m_2-m_3}
.\eea

For simplicity, we assume that there is no primordial $B$ field such as due to a gravitational wave background
and find the following expression for the lensed $E$-field:
\bea \tilde{E}_{lm}=E_{lm}+
\frac{1}{2}\sum_{l_1m_1l_2m_2}\phi_{l_1m_1}E_{l_2m_2}{}_{+2}I_{ll_1l_2}^{mm_1m_2}
(1+(-1)^{l+l_1+l_2})\nonumber\\
+\frac{1}{4}\sum_{l_1m_1l_2m_2l_3m_3}\phi_{l_1m_1}E_{l_2m_2}\phi_{l_3m_3}^*
{}_{+2}J_{ll_1l_2l_3}^{mm_1m_2m_3}(1+(-1)^{l+l_1+l_2+l_3}) \, ,
\label{lensedE}\eea
where the expressions for the mode coupling integrals ${}_{+2}I_{ll_1l_2}^{mm_1m_2}$ and ${}_{+2}J_{ll_1l_2l_3}^{mm_1m_2m_3}$ are
  described in Refs.~\cite{Hu01,CooLi}.

As for the covariance of $E$-mode powerspectrum, we write \bea
\text{Cov}_{\text{EE}} \equiv
\frac{1}{2l_1+1}\frac{1}{2l_2+1}\sum_{m_1m_2}\langle
\tilde{E}_{l_1m_1}\tilde{E}_{l_1m_1}^*\tilde{E}_{l_2m_2}\tilde{E}_{l_2m_2}^*
\rangle
-\tilde{C}_{l_1}^E\tilde{C}_{l_2}^E=\mathcal{H}+\mathcal{I}
+(\mathcal{J}+\mathcal{K})\delta_{l_1l_2}\label{ecov}\eea
where\bea
\mathcal{H}&=&\frac{1}{(2l_1+1)(2l_2+1)}\sum_{L}{C_L^\phi}\left[({}_2F_{l_1Ll_2}C_{l_2}^E)^2
+({}_2F_{l_2Ll_1}C_{l_1}^E)^2\right]
(1+(-1)^{l_1+l_2+L})\nonumber\\
\mathcal{I}&=&\frac{2}{(2l_1+1)(2l_2+1)}\sum_{L}C_L^\phi C_{l_1}^E
C_{l_2}^E(1+(-1)^{l_1+L+l_2}){}_2F_{l_1Ll_2}{}_2F_{l_2Ll_1}\nonumber\\
\mathcal{J}&=&\frac{2}{(2l_1+1)^2}\sum_{L,l'}{C_L^\phi}C_{l'}^EC_{l_1}^E
(1+(-1)^{l_1+L+l'})({}_2F_{l_1Ll'})^2\nonumber\\
\mathcal{K}&=&-\frac{(l_1(l_1+1)-4)}{2\pi(2l_1+1)}\sum_LC_L^\phi(C_{l_1}^E)^2
L(L+1)(2L+1) \, .\eea The last two terms can be written in terms of
the lensed power spectrum of $E$-mode anisotropies as \bea
\mathcal{J}+\mathcal{K}&=&\frac{2}{2l_1+1}(\tilde{C}_{l_1}^E)^2 \,
, \eea
where \bea \tilde{C}_l^E&=&[1-(l^2+l-4)R]C_{l}^E+
\frac{1}{2}\sum_{l_1l_2}C_{l_1}^\phi\frac{({}_2F_{ll_1l_2})^2}{2l+1}
C_{l_2}^E(1+(-1)^{l+l_1+l_2})\nonumber\\
R&=&\frac{1}{8\pi}\sum_{l_1}l_1(l_1+1)(2l_1+1)C_{l_1}^\phi\nonumber \\
{}_2F_{ll_1l_2}&=&\frac{1}{2}[l_1(l_1+1)+l_2(l_2+1)-l(l+1)]\sqrt
{\frac{(2l+1)(2l_1+1)(2l_2+1)}{4\pi}}\begin{pmatrix}l&l_1&l_2\\2&0&-2\end{pmatrix} \, .
\eea
Note that $\tilde{C}_l^E$ is the power spectrum of the lensed $E$-modes.
\begin{figure}[t]
\includegraphics[scale=0.8,angle=0]{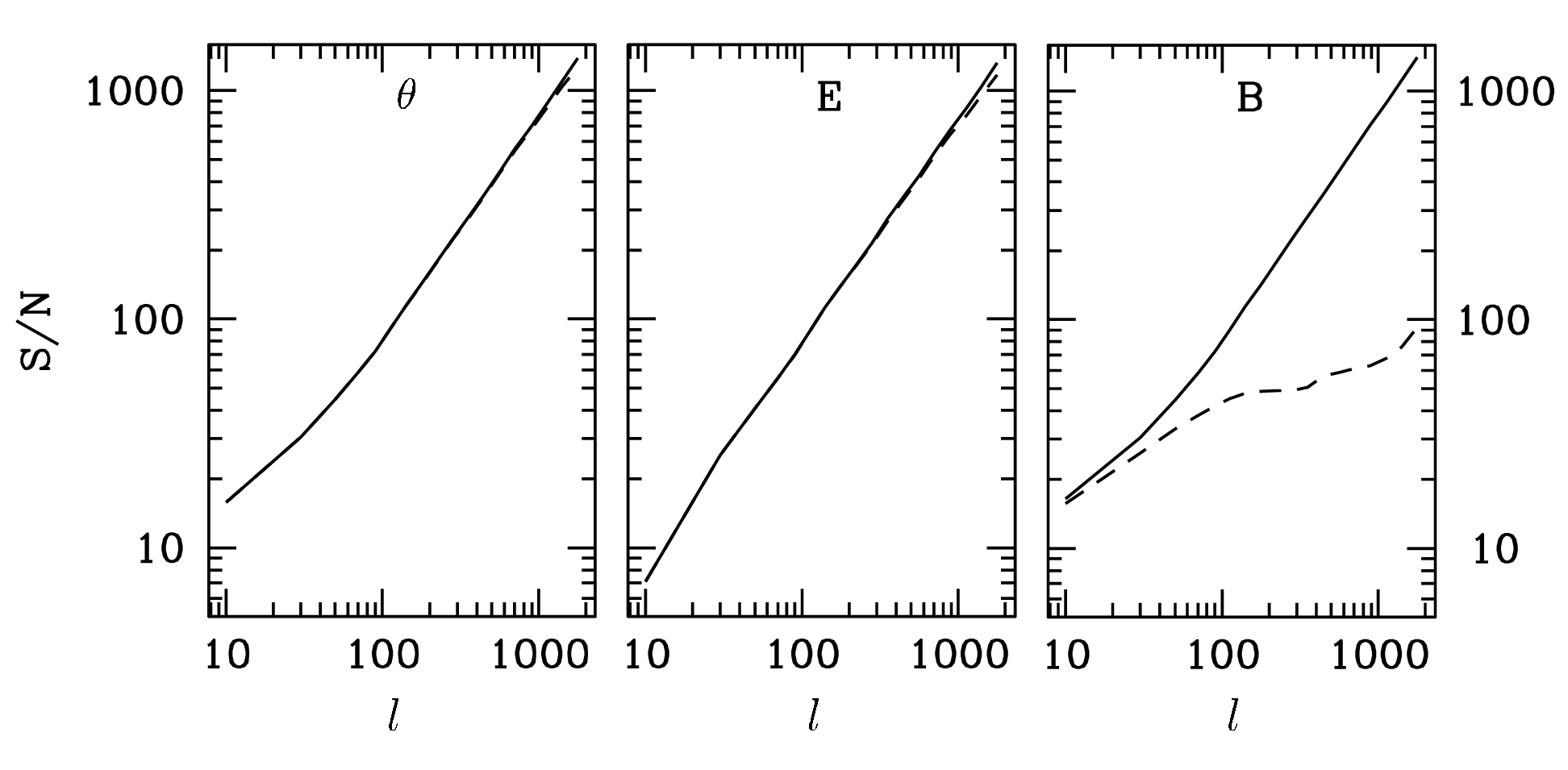}
\caption{ Here we show the cumulative signal-to-noise ratio for a
detection of the power spectrum [Eq.~(\ref{eq:SN})] for
temperature (left), $E$-mode (middle), and $B$-mode (right)
polarization power spectra.  The solid line is the case with a
Gaussian covariance whereas the dashed line is with a non-Gaussian
covariance. We can see that for the case of the temperature and
$E$-mode polarization there is little difference between the
Gaussian and non-Gaussian covariance, but for the $B$-mode
polarization there is a difference of a factor  of $\sim 10$ at
large $l$ values. } \label{sn}
\end{figure}

\subsection{$B$-mode Polarization Covariance}

The calculation related to $B$-mode power spectrum polarization is similar to the case of the $E$-modes
except that we assume that the $B$-mode polarization is generated solely the lensing of the $E$-mode polarization.
Based on previous work (c.f., Ref.~\cite{Hu00}), we write the multipole moments of the lensed $B$-modes as
\bea i\tilde{B}_{lm} &=&
\frac{1}{2}\sum_{l_1m_1l_2m_2}\phi_{l_1m_1}E_{l_2m_2}{}_{+2}I_{ll_1l_2}^{mm_1m_2}
(1-(-1)^{l+l_1+l_2})\nonumber\\
&+&\frac{1}{4}\sum_{l_1m_1l_2m_2l_3m_3}\phi_{l_1m_1}E_{l_2m_2}\phi_{l_3m_3}^*
{}_{+2}J_{ll_1l_2l_3}^{mm_1m_2m_3}(1-(-1)^{l+l_1+l_2+l_3}).
\label{lensedB}\eea
\begin{figure}[t]
\includegraphics[scale=0.8,angle=0]{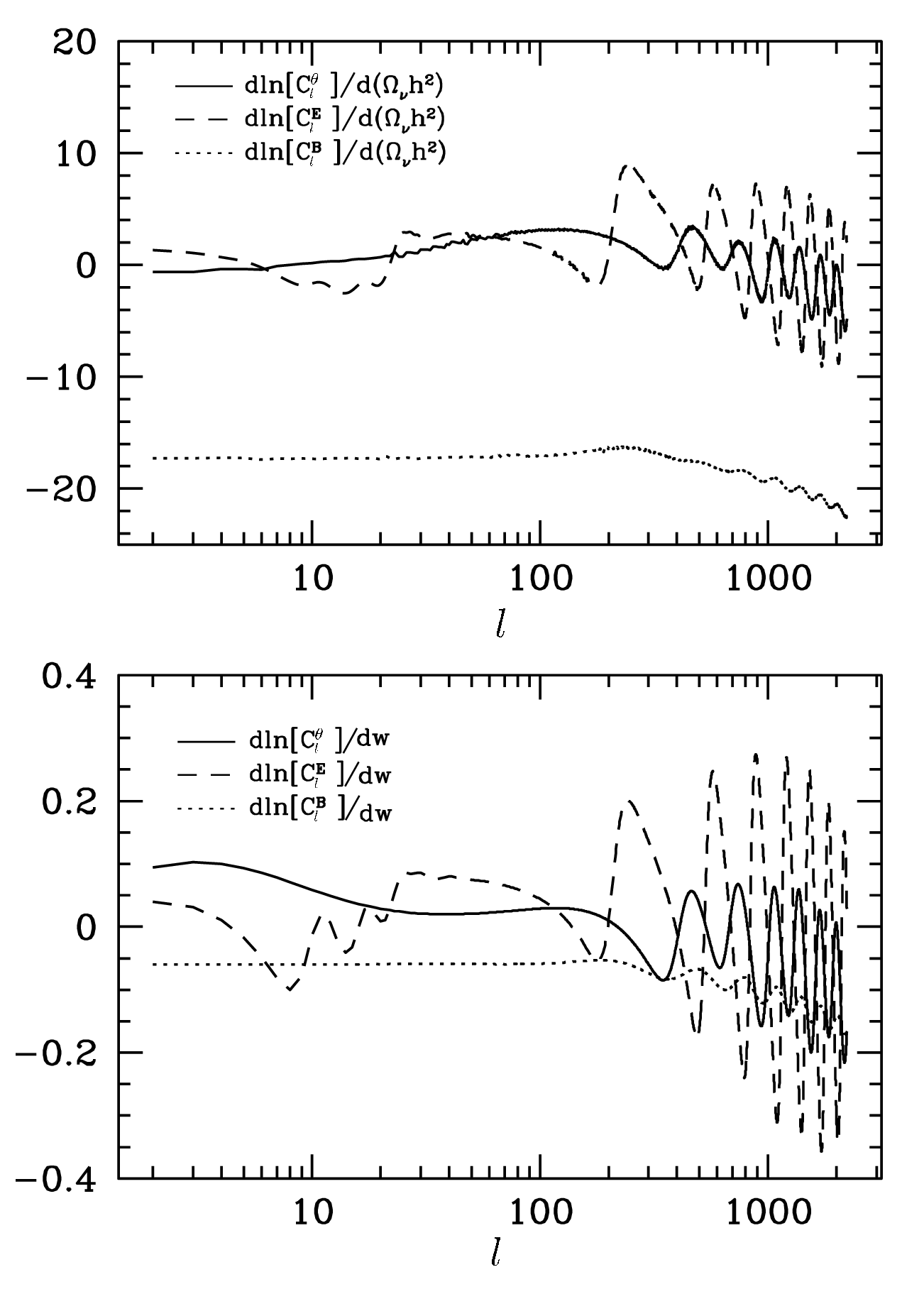}
\caption{The derivatives of the temperature ($\theta$), $E$-mode, and $B$-mode power spectra with respect to the sum of the neutrino masses ($\propto \Omega_{\nu} h^2$, top
panel) and
the dark energy equation of state, $w$ (bottom panel).  It is clear that in the case of the sum of the neutrino
masses the addition of the $B$-mode polarization greatly increases sensitivity.  In both cases we find that large $l$ information
also increases sensitivity.  We note that the derivative of the temperature power spectrum with respect to neutrino mass agrees with
that shown in Fig.~3 of Ref.~\cite{Eisenstein:1998hr}.
}
\label{spectra}
\end{figure}
Here, we will only calculate the $B$-mode trispectrum with terms involving $C_l^\phi$ since
we will make the assumption that corrections to $B$-modes from the bsipectrum and higher-order
non-Gaussianities of the lensing $\phi$ field are subdominant. Thus, using the first term of the expansion,
we write
\bea
&&\langle\tilde{B}_{l_1m_1}\tilde{B}_{l_2m_2}\tilde{B}_{l_3m_3}\tilde{B}_{l_4m_4}
\rangle=\frac{1}{16}\sum_{L_1M_1l'_1m'_1} \sum_{L_2M_2l'_2m'_2}
\sum_{L_3M_3l'_3m'_3}\sum_{L_4M_4l'_4m'_4}\nonumber\\
&&\langle\phi_{L_1M_1}\phi_{L_2M_2}\phi_{L_3M_3}\phi_{L_4M_4}
\rangle \langle
E_{l'_1m'_1}E_{l'_2m'_2}E_{l'_3m'_3}E_{l'_4m'_4}\rangle
{}_{+2}I_{l_1m_1L_1M_1l'_1m'_1}{}_{+2}I_{l_2m_2L_2M_2l'_2m'_2}{}_{+2}I_{l_3m_3L_3M_3l'_3m'_3}
{}_{+2}I_{l_4m_4L_4M_4l'_4m'_4}\nonumber\\
&&(1-(-1)^{l_1+L_1+l'_1})(1-(-1)^{l_2+L_2+l'_2})(1-(-1)^{l_3+L_3+l'_3})
(1-(-1)^{l_4+L_4+l'_4})\nonumber\\
\eea where \bea
{}_{+2}I_{lml_1m_1l_2m_2}&=&{}_2F_{ll_1l_2}(-1)^m\begin{pmatrix}
l&l_1&l_2\\-m&m_1&m_2 \, .
\end{pmatrix}
\eea
\begin{figure*}[t]
\includegraphics[scale=0.6,angle=0]{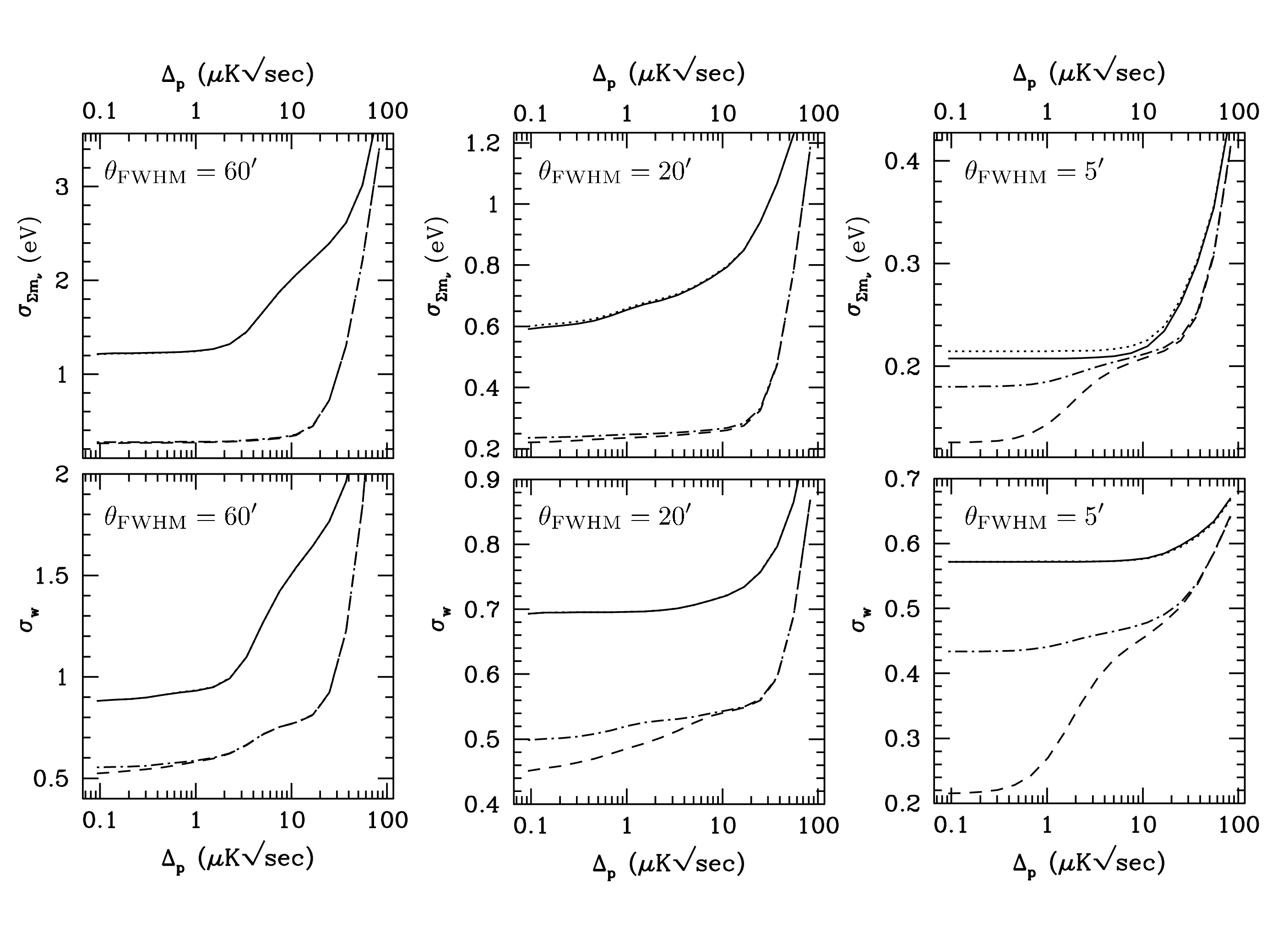}
\caption{
 The expected error on the sum  of the neutrino masses
(top three panels) and the dark energy equation of state, $w$
(bottom three panels) as a function of experimental noise for
three different values of the beam width,
$\theta_{\mathrm{FWHM}}$.  The solid line considers Gaussian
covariance with just temperature information, the dotted line
considers non-Gaussian covariance with just temperature
information, the dashed line considers Gaussian covariance with
both temperature and polarization ($E$- and $B$-mode), and the
dot-dashed line considers non-Gaussian covariance with both
temperature and polarization.  It is clear that as the beam width
is decreased the estimated error on the sum of the neutrino masses
and $w$ is increasingly overly optimistic when just the Gaussian
covariance is used in the Fisher matrix calculation.  We choose 5
bins uniformly spacing between $l=5$ and $l=100$, while we choose
13 bins logarithmic uniformly spacing between $l=100$ and
$l=2000$. This choice of bins are sparser compared to
\cite{smith06}. From the expressions of covariance matrix
[Eqs.\eqref{thetacov},\eqref{ecov},\eqref{bcov}], we know the
Gaussian parts are diagonal and therefore the larger the bin is,
the more important the non-Gaussian effect is. So the non-Gaussian
effects in our bandpower statistics are more obvious that those in
\cite{smith06}. } \label{noise}
\end{figure*}

 The covariance of the $B$-mode angular power
spectrum can be now defined as \bea \text{Cov}_{\text{BB}} \equiv
\frac{1}{2l_1+1}\frac{1}{2l_2+1}\sum_{m_1m_2}\langle
\tilde{B}_{l_1m_1}\tilde{B}_{l_1m_1}^*\tilde{B}_{l_2m_2}\tilde{B}_{l_2m_2}^*
\rangle-\tilde{C}_{l_1}^B\tilde{C}_{l_2}^B\, . \eea After some
straightforward but tedious algebra, we obtain \bea
\text{Cov}_{\text{BB}}
=\mathcal{A}+\mathcal{B}+\mathcal{C}+\delta_{l_1l_2}\mathcal{D},\label{bcov}\eea
where  the terms are given by \bea
\mathcal{A}=\frac{2}{4(2l_1+1)(2l_2+1)}\sum_{L=1}^{N_\phi}\left[
\frac{(C_{L}^\phi)^2}{2L+1}\left(\sum_{l'=|l_1-L|}^{l_1+L}C_{l'}^E(1-(-1)^{l_1+L+l'})
({}_2F_{l_1Ll'})^2\right)
\left(\sum_{l'=|l_2-L|}^{l_2+L}C_{l'}^E(1-(-1)^{l_2+L+l'})({}_2F_{l_2Ll'})^2\right)\right]\nonumber
\eea \bea
\mathcal{B}&=&\frac{2}{4(2l_1+1)(2l_2+1)}\sum_{l'=1}^{N_E}\left[\frac{(C_{l'}^E)^2}{2l'+1}
\left(\sum_{L=|l_1-l'|}^{l_1+l'} C_L^\phi
(1-(-1)^{l_1+L+l'})({}_2F_{l_1Ll'})^2\right)\left(\sum_{L=|l_2-l'|}^{l_2+l'}
C_L^\phi
(1-(-1)^{l_2+L+l'})({}_2F_{l_2Ll'})^2\right)\right]\nonumber \eea
\bea
\mathcal{C}&=&\frac{2}{16(2l_1+1)(2l_2+1)}\sum_{L_1=1}^{N_\phi}\sum_{l'_1=|l_1-L_1|}^{l_1+L_1}
\sum_{l'_2=|l_2-L_1|}^{l_2+L_1}\sum_{L_2=|l_1-l'_2|}^{l_1+l'_2}C_{L_1}^\phi
C_{L_2}^\phi C_{l'_1}^E
C_{l'_2}^E\left[{}_2F_{l_1L_1l'_1}{}_2F_{l_1L_2l'_2}{}_2F_{l_2L_1l'_2}{}_2F_{l_2L_2l'_1}
\right]\left\{
\begin{aligned}L_1\quad l'_1\quad l_1\\
L_2\quad l'_2\quad l_2\end{aligned}\right\} \nonumber\\
&&(-1)^{l_1+L_1+l'_1+l_2+L_2+l'_2}(1-(-1)^{l_1+L_1+l'_1})(1-(-1)^{l_1+L_2+l'_2})
(1-(-1)^{l_2+L_1+l'_2})(1-(-1)^{l_2+L_2+l'_1}) ,\nonumber\\
\mathcal{D}&=&\frac{2}{4(2l_1+1)^3} \left( \sum_{Ll'}C_L^\phi
C_{l'}^E
[1-(-1)^{l_1+L+l'}]({}_2F_{l_1Ll'})^2\right)^2=\frac{2}{2l_1+1}(\tilde{C}_{l_1}^B)^2,\eea
where \bea \tilde{C}_l^B&=&
\frac{1}{2}\sum_{l_1l_2}C_{l_1}^\phi\frac{({}_2F_{ll_1l_2})^2}{2l+1}
C_{l_2}^E(1-(-1)^{l+l_1+l_2}).\eea

Unlike the calculation for the covariances of the lensed
temperature and polarization $E$-mode, the numerical calculation
related to covariance of the $B$-modes is complicated due to the
term $\mathcal{C}$, which involves a Wigner-6j symbol. These
symbols can be generated using the recursion relation outlines in
Appendix of Ref.~\cite{Hu01}, though we found that such recursions
are subject to numerical instabilities when one of the $l$ values
is largely different from the others  and the $l$ values are
large. In these cases, we found that values accurate to better
than a ten percent of the exact result can be obtained through
semi-classical formulae \cite{Shulten}. In any case, we found that
$\mathcal{C}$ is no more than $1\%$ of  $\mathcal{A}$,
$\mathcal{B}$, and these terms are in turn no more than $10\%$ of
$\mathcal{D}$. The same situation happens to those expressions in
flat sky approach\cite{smith06}.

\begin{figure}[t]
\includegraphics[scale=0.75,angle=0]{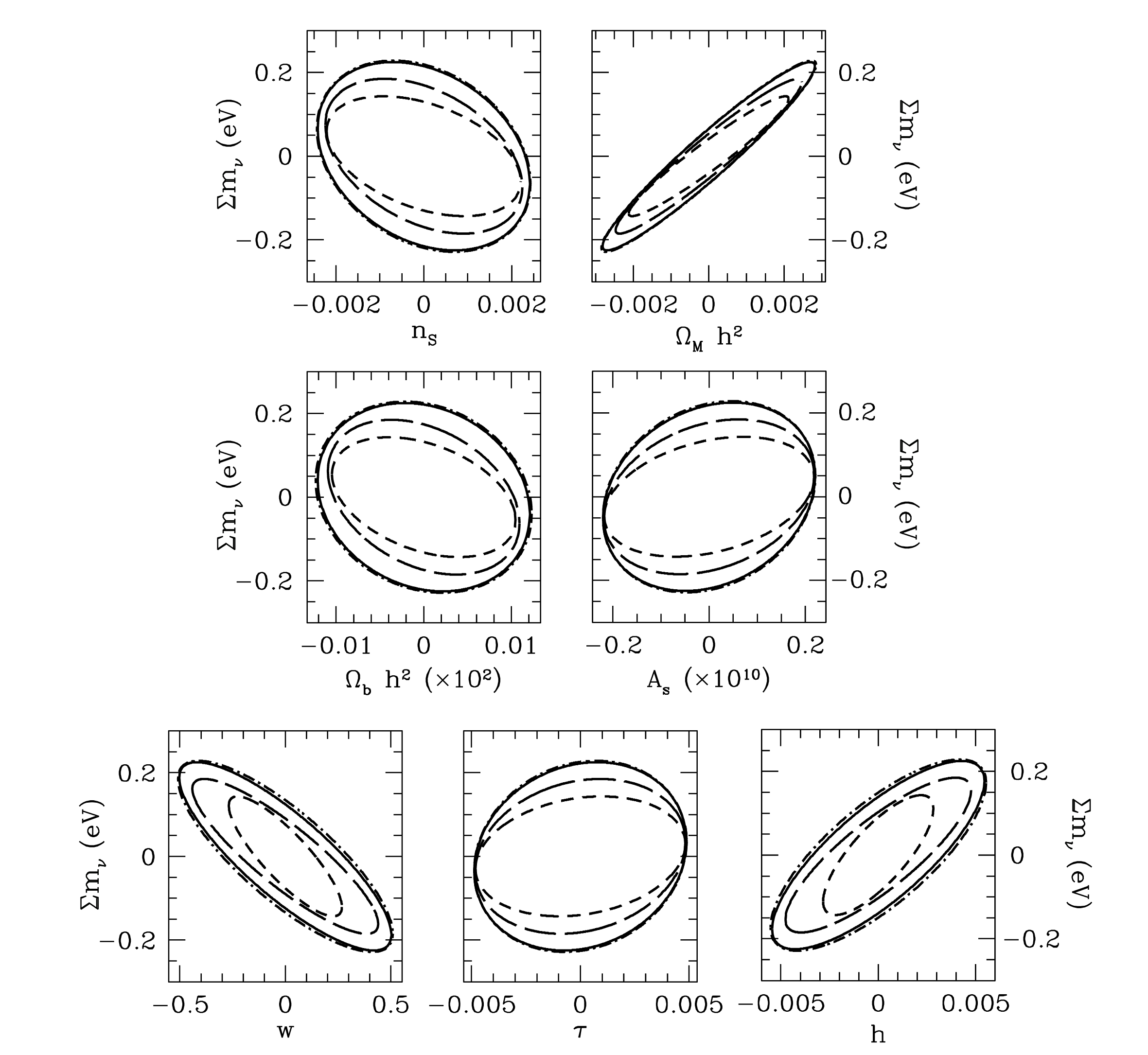}
\caption{
The error ellipses from our Fisher matrix calculation.  We have varied eight parameters, and show the error ellipses for each parameter with $\Sigma m_{\nu}$.  The dot-dashed ellipse is the expected error from Planck with just a Gaussian covariance, the solid ellipse is same but with a non-Gaussian covariance.  The short-dashed ellipse is for an experiment with the same beam width as Planck ($\theta \sim 5^{\prime}$) but with decreased noise ($1\ \mu\mathrm{K} \sqrt{\mathrm{sec}}$ as opposed to $25\ \mu\mathrm{K} \sqrt{\mathrm{sec}}$) with a gaussian covariance and the long-dashed ellipse is the same but with a non-Gaussian covariance.
}
\label{ellipses}
\end{figure}

\section{Results and Discussion}
\label{sec:results}

We begin our discussion on the parameter
uncertainties in the presence of non-Gaussian covariance by first
establishing that one cannot ignore them for the $B$-mode power sptectrum.
In Figure~1 we show the correlation matrix, which is defined as
\begin{equation}
r_{ij} \equiv \frac{{\rm
Cov}_{XY}(i,j)}{\sqrt{{\rm Cov}_{XY}(i,i){\rm Cov}_{XY}(j,j)}}.
\label{eq:correlation}
\end{equation}
This correlation normalizes the diagonal to unity and displays the
off diagonal terms as a value between 0 and 1. This facilitates an
easy comparison on the importance of non-Gaussianities between
temperature, $E$-, and $B$-modes of polarization. As shown in
Figure~1, the off diagonal entries of temperature and $E$-modes are
roughly at the level of $10^{-5}$ suggesting that non-Gaussian
covariance is not a concern for these observations out to
multipoles of 2000 \cite{Coo02}, while for $B$-modes the
correlations are at the level above 0.1 and are significant.

Below when we calculate the signal to noise ratio and Fisher
matrices, we use the bandpowers as observables with logrithmic
bins in the multipole space. Our bandpower estimator for two
quantities of X- and Y-fields involving temperature and
polarization maps is \bea
\hat{\Delta}_{XY,i}^2&=&\frac{1}{\alpha_i}\sum_{l=l_{i1}}^{l_{i2}}\sum_{m=-l}^{l}
\frac{l}{4\pi}X_{lm}Y_{lm}^* \eea where  $\alpha_i=l_{i2}-l_{i1}$
is an overall normalization factor given by the bin width. The
angular power spectra are \bea {\Delta}_i^2=\langle
\hat{\Delta}_i^2\rangle=\frac{1}{4\pi
\alpha_i}\sum_{l}(2l+1)lC_l^{B,E,\theta} \, , \eea while the full
covariance matrix is \bea \langle
(\hat{\Delta}_i^2-\Delta_i^2)(\hat{\Delta}_j^2-\Delta_j^2)\rangle=S_{ii}^G\delta_{ij}+S_{ij}^N
\, , \eea with the Gaussian part \bea
S_{ii}^G&=&\frac{2}{(4\pi)^2\alpha_i^2}\sum_{l_1=l_{i1}}^{l_1=l_{i2}}
(2l_1+1)l_1^2(C_{l_1}^{B,E,\theta}+N_{l_1})^2,\nonumber
\\ N_l&=&\left( \frac{\Delta_p}{T_{\rm CMB}}\right)^2 e^{l(l+1)\theta^2_{\text{FWHM}}/{8\ln 2}}
\label{noise}\eea  and the non-Gaussian part is \bea
S_{ij}^N=\frac{1}{(4\pi)^2
\alpha_i\alpha_j}\sum_{l_1l_2}(2l_1+1)(2l_2+1)l_1l_2
(\text{Cov}_{B,E,\theta}^N) \, . \eea

To further quantify the importance of non-Gaussianities for
$B$-modes, in Figure~2, we plot the cumulative signal-to-noise ratio
for the detection of the power spectra as a function of the
bandpowers. These are calculated as
\begin{equation}
\left({\rm S}{\rm N}\right)_{XY} = \sum_{\Delta_i \Delta_j}
C_{\Delta_i}^{XY} {\rm Cov}^{-1}_{XY}(\Delta_i,\Delta_j)
C_{\Delta_j}^{XY} \, ,
\label{eq:SN}
\end{equation}
by ignoring the instrumental noise contribution to the covariance.
As shown, there is no difference in
the signal-to-noise ratio for the temperature and  $E$-mode power
spectra measurement due to non-Gaussian covariances, while there
is a sharp reduction in the cumulative signal-to-noise ratio for a
detection of the $B$-modes. This reduction is significant and can be
explained through the effective reduction in the number of
independent modes at each multipole from which clustering
measurements can be made. In the case of Gaussian statistics, at
each multipole $l$, there are $2l+1$ modes to make the power
spectrum measurements. In the case of non-Gaussian statistics with
a covariance, this number is reduced further by the correlations
between different modes. If $N$ is the number of independent modes
available under Gaussian statistics, a simple calculation shows
that the effective number of modes are reduced by $[1+(N-1)r^2]$
when the modes are correlated by an equally distributed
correlation coefficient $r$ among all modes. With $N=2l+1$ and
substituting a typical correlation coefficient $r$ of 0.15, we
find that the cumulative signal-to-noise
 ratio should be reduced by a factor of 7 to 8 when compared to the case where only Gaussian statistics are assumed. This is
consistent with the signal-to-noise ratio estimates shown in Figure~2 based on an exact calculation using the full covariance
matrix that suggests a slightly larger reduction due to the fact that some of the modes are more strongly correlated than the assumed
average value.

To calculate the overall impact on cosmological parameter
measurements using temperature and polarization spectra, we make
use of the Fisher information matrix given for tow parameters
$\mu$ and $\nu$ as \bea
F_{\mu\nu}=\sum_{X=B,E,\theta}\sum_{ij}\frac{\partial
(\Delta_i^X)^2}{\partial
p_\mu}(\text{Cov}_{XX}^{-1})\frac{\partial
(\Delta_j^{X})^2}{\partial p_\nu},\eea where the summation is over
all bins. While this is the full Fisher information matrix, we
will divide our results to with and without non-Gaussian
covariance as well as to information on parameters present within
temperature, and $E$- and $B$-modes of polarization.

Since $B$-modes have been generally described as a probe of neutrino mass and the dark energy equation of state, in Figure~3,
we show $\partial C_l/\partial m_\nu$ and $\partial C_l/\partial w$  to show the extent to which information on these two quantities
are present in the spectra. It is clear that $B$-modes are a strong probe of neutrino mass given that the sensitivity of temperature
and $E$-modes are smaller compared to the fractional difference in the $B$-modes. Furthermore, $B$-modes also have some senitivity to the
dark energy equation of state, but fractionally, this sensitivity is smaller compared to the information related to the
neutrino mass.

In Figure~4, we summarize parameter constraints on these two
parameters as a function of the instrumental noise for different
values of resolution with and without non-Gaussian covariance.
While for low resolution experiments the difference between
Gaussian and non-Gaussian extraction is marginal,
non-Gaussianities become more important for high resolution
experiments where one probes $B$-modes down to large multipoles. In
this case, the parameters  extraction is degraded by up to a
factor of more than 2.5 for both the neutrino mass and the dark
energy equation of state. We have not attempted to calculate the
parameter errors for experiments with resolution better than 5
arcminutes. This is due to the fact that such experiments will
probe multipoles higher than 2000 and we are concerned that we do
not have a full description of the non-Gaussian covariance at such
small scales due to uncertainties in the description of the matter
power spectrum at non-linear scales. As described in
Ref.~\cite{Cha}, the CMB lensing calculation must account for
non-linearities and their importance only become significant for
small angular scale anisotropy experiments. Furthermore, we also
do not think any of the upcoming $B$-mode polarization experiments
with high sensitivity, which will be either space-based or
balloon-borne, will have large apertures to probe multipoles above
2000.

The value of 2000 where we stop our calculations is also
consistent with Planck. Since Planck HFI experiment will have a
total focal plane polarization noise of about 25 $\mu$K $\sqrt{\rm
sec}$, based on Figure~4, we find that it will constrain the
neutrino mass to be below 0.22 eV and the dark energy equation of
state will be determined to an accuracy of 0.5. Note that the
combination  of Planck noise and resolution is such that one does
not find a large difference between Gaussian and non-Gaussian
statistics, but on the otherhand, experiments that improve the
polarization noise well beyond Planck must account for
non-Gaussian noise properly. In future, there are plans for a
Inflation Probe or a CMBpol mission that will make high sensitive
observations in search for a gravitational wave background. If
such an experiment reach an effective  noise level of 1 $\mu$K
$\sqrt{\rm sec}$ and has the same resolution as Planck, the
combined polarization observations can constrain the neutrino mass
to be about 0.18 while the dark energy equation of state will be
known to an accuracy of 0.44. This is well above the suggested
constraint from Gaussian noise level. This suggests that while
high sensitive  $B$-mode measurements are desirable for studies
involving the gravitational wave background, they are unlikely to
be helpful for increasingly better constraints on the cosmological
parameters.

The non-Gaussianities in the $B$-modes, while providing information on gravitational lensing,
limits accurate parameter estimates from the power spectrum alone. This is contrary to some of the suggestions
in the literarture that have indicated high precision of measurements on parameters such as the neutrino mass
and the dark energy equation of state with CMB $B$-mode powerspectrum by ignoring issues related to non-Gaussian
correlations. Furthermore, while atmospheric oscillations suggest a mass-squared difference of $\Delta m_\nu^2 \sim 10^{-3}$
for two of the neutrino species, it is unlikely that one will be able to distinguish between
mass hierarchies with CMB polarization observations alone if one of the two masses related to the atmospheric
oscillation result is close to zero (c.f. \cite{Les06}). This is discouraging, but understanding
the information present in CMB polarization beyond powerspectra, such as direct measurements of non-Gaussianities
themselves, could potentially allow an improvement.

From Figure 4 we see that as we decrease $\Delta_p$ the measurement errors on the parameters
asymptote to a constant value.  We can understand this in the following way.  As we see from
Eq.~\eqref{noise}, the noise blows up exponentially at large $l$
and therefore sets an effective cutoff $l_0$.  Only the bandpowers
which are smaller than $l_0$ contribute to parameter estimates.
Therefore, if we decrease $\Delta_p$, we increase
the number of bandpowers we can observe and
hence obtain better sensitivity with negligible instrumental noise for $l \lesssim l_0$.
Therefore, the curves in Figure 4 become flatter as we decrease
$\Delta_p$. The same situation applies to Figure 5.  Figure 4 also shows that as we decrease the beam width,
$\theta_{\mathrm{FWHM}}$, we see the Gaussian covariance becomes more significant.  This is a
result of the fact that the Gaussian covariance grows in significance with increasing $l$.

In Figure~5, to highlight the impact on cosmological parameters
beyond the neutrino mass and dark energy equation of state, we
also show constraints from the Fisher matrix calculation. We show
error ellipses calculated with and without the non-Gaussian
lensing covariance for two different experiments: Planck, with
$\theta_{\mathrm{FWHM}} = 5^{\prime}$ and $\Delta_p = 25\
\mu\mathrm{K} \sqrt{\mathrm{sec}}$ and `super-Planck' with
$\theta_{\mathrm{FWHM}} = 5^{\prime}$ and $\Delta_p = 1\
\mu\mathrm{K} \sqrt{\mathrm{sec}}$. This comparison shows that
while parameters such as $m_\nu$ and $w$ are affected, parameters
such as $\tau$, $\Omega_mh^2$ are not affected by non-Gaussian
information. This is due to the fact that the cosmological
information on these parameters come from temperature and
$E$-modes rather than $B$-modes. This highlights the fact that the
issues discussed here are primarily a concern for the $B$-mode
measurements and extraction of parameters, especially the
parameters that have been recognized to be mostly constrained by
the $B$-mode measurements, and not for temperature and $E$-modes.

\section{Summary}
\label{sec:summary}

The $B$-mode polarization lensing signal is a useful probe of certain cosmological
parameters such as the neutrino mass and the dark energy equation of state as the signal depends on the
integrated mass power spectrum between us and the last scattering surface. This lensing $B$-mode
signal, however, is non-Gaussian and the resulting non-Gaussian covariance to the powerspectrum cannot
be ignored when compared to the case of temperature and polarization $E$-mode anisotropy covariances.
The resulting degradation on neutrino mass and dark energy equation of state is about a factor of 2
when compared to the case where statistics are simply considered to be Gaussian.
We discuss parameter uncertainties achievable in upcoming experiments.

\begin{acknowledgments}
During the preparation of the paper, we learned of another completed work involving lensing non-Gaussian
covariance by Smith et al. \cite{smith06} that also reached conclusions similar to ours on parameter measurements.
We thank Wayne Hu, Manoj Kaplinghat, and Kendrick Smith for useful discussions and communications.
This work was supported in part by DoE at UC Irvine (AC), by the Moore Foundation at Caltech (CL), and a NSF graduate research fellowship (TLS).
\end{acknowledgments}


\end{document}